\documentclass[11pt]{article}
\usepackage{tabularx} 
\usepackage{graphicx}
\usepackage[graphicx]{realboxes}

\usepackage{adjustbox}
\usepackage{color,xcolor}
\usepackage{colortbl}
\usepackage{array}
\usepackage{bigstrut}

\usepackage{multirow}
\usepackage{mathtools}
\usepackage{morefloats}
\usepackage{amsfonts}
\usepackage{color}

\usepackage{latexsym}
\usepackage{amsmath,amssymb,amsthm}
\usepackage{dsfont}
\usepackage{extarrows}
\usepackage{hyperref}
\usepackage{lscape}
\usepackage{units}
\usepackage{amsmath}
\usepackage{natbib}

\usepackage{tikz}
\newcommand*\circled[1]{\tikz[baseline=(char.base)]{
            \node[shape=circle,draw,inner sep=2pt] (char) {#1};}}

\newcommand*\circledfill[1]{\tikz[baseline=(char.base)]{
            \node[shape=circle,fill=black!25,draw,inner sep=2pt] (char) {#1};}}

\newcommand*\squared[1]{\tikz[baseline=(char.base)]{
            \node[shape=rectangle,draw,inner sep=2pt] (char) {#1};}}

\newcommand*\squaredfill[1]{\tikz[baseline=(char.base)]{
            \node[shape=rectangle,fill=black!25,draw,inner sep=2pt] (char) {#1};}}

\usepackage{circledtext}

\numberwithin{equation}{section}

\newtheoremstyle{thm}
{9pt}
{9pt}
{\itshape}
{}
{\bfseries}
{.}
{ }
{}
\theoremstyle{thm}

\newtheoremstyle{def}
{9pt}
{9pt}
{}
{}
{\bfseries}
{.}
{ }
{}
\theoremstyle{def}

\renewcommand{\footnoterule}{%
 & \kern -3.5pt
 & \hrule width \textwidth height 1pt
 & \kern 3.5pt
}

\makeatletter
\def\blfootnote{\xdef\@thefnmark{}\@footnotetext}
\makeatother

\begin{document}

\title{\bf Revisiting the memoryless property -- testing for the Pareto type I distribution}


\author{L. Ndwandwe, J.S. Allison, L. Santana and I.J.H. Visagie\\
Pure and Applied Analytics, North-West University, South Africa\\
lethani.ndwandwe@nwu.ac.za}

\date{}
\maketitle

\begin{abstract}
We propose new goodness-of-fit tests for the Pareto type I distribution. These tests are based on a multiplicative version of the memoryless property which characterises this distribution. We present the results of a Monte Carlo power study demonstrating that the proposed tests are powerful compared to existing tests.
%
As a result of independent interest, we demonstrate that tests specifically developed for the Pareto type I distribution substantially outperform tests for exponentiality applied to log-transformed data (since Pareto type I distributed values can be transformed to exponentiality via a simple log-transformation).
Specifically, the newly proposed tests based on the multiplicative memoryless property of the Pareto distribution substantially outperform a test based on the memoryless property of the exponential distribution.
%
The practical use of tests is illustrated by testing the hypothesis that two sets of observed golfers' earnings (those of the PGA and LIV tours) are realised from Pareto distributions.
\end{abstract}

\textbf{Key words:} Memoryless property, Goodness-of-fit testing, Pareto distribution.

\section{Introduction}

The memoryless property of the exponential distribution is well-known and characterises 
this distribution among the continuous laws. 
This property implies that 
if, for example, the exponential distribution is used to model waiting times,
then the probability of an event occurring within the next unit of time is independent 
of the amount of time that has passed. 
Let $Y$ be a random variable denoting an exponential waiting time. 
The memoryless property of the exponential distribution can be expressed as
\begin{equation}
    P(Y \ge s+t|Y>s)=P(Y>t), \quad t,s\ge 0. \label{a}
\end{equation}
For more details regarding this and some other characterisations of the exponential distribution, see \cite{tomy2020review} as well as \cite{galambos2006characterizations}. 
The memoryless property has been used to develop several goodness-of-fit tests for the exponential distribution, see \cite{ahmad1999goodness} as well as \cite{alwasel2001goodness}. Below, we use a related characterisation to propose new goodness-of-fit tests for the Pareto distribution.

The Pareto distribution is named after the Italian economist Vilfredo Pareto, who first described this law in his 1897 book \emph{Cours d'\'{e}conomie politique}, see \cite{Par}. Several generalisations of this distribution have since been proposed, see \cite{Arn}, and the originally proposed distribution has subsequently become known as the Pareto type I distribution. In what follows, we refer to the Pareto type I distribution simply as the Pareto distribution. This distribution is closely related to the exponential distribution. If $Y$ is exponential with mean $1/\beta$, then $X=\textrm{e}^Y$ is a Pareto distributed random variable with shape parameter $\beta>0$, denoted by $X\sim P(\beta)$. In this case, the distribution function of $X$ is
\begin{equation*}
    F(x)=1-x^{-\beta}, \quad \beta>0, \quad x>1.
\end{equation*}
Note that the Pareto distribution is often specified to include a scale parameter; we take this parameter to be equal to 1 throughout. Similar results to those presented in this paper can be obtained when the scale parameter is required to be estimated from observed data.

The Pareto distribution is widely used to model phenomena exhibiting heavy tails. Examples of the use of the Pareto distribution include the modelling of excess losses in insurance claims, see \cite{rytgaard1990estimation}, as well as the modelling of failure times of mechanical components, see \cite{BS:2018}. The Pareto is also a popular model found in survival analysis and reliability theory, see, \cite{NAV2021} as well as \cite{nofal2017new}. For a recent review of goodness-of-fit tests for the Pareto distribution, see \cite{ndwandwe2022testing} and for more general Pareto distributions see \cite{chu2019}. 

The link between the exponential and Pareto laws means that the memoryless property of the exponential distribution has implications for the Pareto distribution as well. The Pareto distribution exhibits a multiplicative version of the memoryless property, which can be expressed as follows: $X \sim P(\beta)$ for some $\beta>0$ if, and only if,
\begin{equation}
     P(X \ge st\,|\,X>s)=P(X>t), \quad t,s\ge 1.\label{b}
\end{equation}  
This property will be exploited to construct goodness-of-fit tests for the Pareto distribution. Naturally, due to the link between the Pareto and exponential distributions, tests for exponentially can also be applied to log-transformed data in order to test for the Pareto distribution.

The remainder of the paper is structured as follows. In Section \ref{sect2}, we introduce two new classes of goodness-of-fit tests based on a characterization that uses the multiplicative memoryless property. Section \ref{sect3} includes a Monte Carlo study comparing the performances of the newly proposed tests to those of existing tests for the Pareto distribution as well as  tests for exponentiality applied to transformed data. The finite sample performances are considered against a wide range of fixed alternative distributions. Additionally, the performances of the tests against a range of mixture distributions are also compared in this section. In Section \ref{sect4}, we demonstrate the use of the tests in order to determine whether or not observed data sets, relating to the earning of golfers, are realised from a Pareto distribution. Finally, Section \ref{sect5} presents some conclusions.


\section{The test statistics based on the memoryless property} \label{sect2}

In this section we develop two classes of goodness-of-fit tests for the Pareto distribution based on the characterisation given in \eqref{b}. Given a random sample $X, X_1, X_2,..., X_n$, from an unknown probability distribution $F$ with density $f$, we wish to test the composite null hypothesis
\begin{equation}\label{hyp}
  H_0: X\sim P(\beta),
\end{equation}
for some $\beta>0$, against general alternatives.

To estimate $\beta$, we use two different estimators, namely, the maximum likelihood estimator (MLE) and the method of moments estimator (MME).
The MLE is
\begin{equation}\label{betamle}
    \widetilde{\beta}_n: = \widetilde{\beta}(X_1,...,X_n) = \frac{n}{\sum_{j=1}^{n}\log (X_j)},
\end{equation}
while the MME is
\begin{equation*}
    \ddot{\beta}_n:=\ddot{\beta}(X_1,...,X_n) =   \frac{\Bar{X}}{\Bar{X}-1},
\end{equation*}
with $\Bar{X} = \frac{1}{n} \sum_{j=1}^n X_j.$ Henceforth, we will use the notation $\widehat\beta_n$ to denote either of these two estimators.


Upon setting $t=s$ in \eqref{b}, we obtain that $X \sim P(\beta)$ if, and only if,
\begin{equation}
    P(X>t^2)=[P(X>t)]^2, \quad t\ge 1. \label{c}
\end{equation}
We now propose a test statistic based on the characterisation given in \eqref{c}. By estimating the left hand side of \eqref{c} nonparametrically using the empirical survival function, $S_n(t)=\frac{1}{n}\sum_{j=1}^{n}I(X_j>t)$, and the right hand side parametrically by using the survival function of the Pareto distribution with estimated $\widehat{\beta}_n$, $\bigl[S_{\widehat{\beta}_n}(t)\bigl]^2=t^{-2\widehat{\beta}_n}$, we suggest the test statistic
\begin{align*}
      MP^{(1)}_n& =\int_{1}^{\infty}\left(S_n(t^2)-\bigl[S_{\widehat{\beta}_n}(t)\bigl]^2\right)^2 \text{d}F_{\widehat{\beta}_n}(t)\\
      & =\int_{1}^{\infty}\left(\frac{1}{n}\sum_{j=1}^{n}I(X_{j}>t^2)-t^{-2\widehat{\beta}_n}\right)^2\widehat{\beta}_nt^{-(\widehat{\beta}_n+1)}\text{d}t.
\end{align*}
From some straightforward calculation we have that the test statistic simplifies to
\begin{equation*}
    MP^{(1)}_n=\frac{2}{3n}\sum_{j=1}^{n}{X_j}^{-3\widehat{\beta}_n/2}-\frac{1}{n^2}\sum_{j=1}^{n}\sum_{k=1}^{n}\min({X_j},{X_k})^{-\widehat{\beta}_n/2}+\frac{8}{15}.
\end{equation*}
The double summation in the expression for $MP^{(1)}_n$ can be reduced to a single summation based on the order statistics, $X_{(1)} < X_{(2)} < \cdots < X_{(n)}$,
\begin{equation*}
    MP^{(1)}_n=\frac{2}{3n}\sum_{j=1}^{n}{X_j}^{-3\widehat{\beta}_n/2}- \frac{1}{n^2}\sum_{j=1}^{n} v_{j,n} X_{(j)}^{-\widehat{\beta}_n/2}+\frac{8}{15},
\end{equation*}
where $v_{j,n}:= (n-j+1)^2-(n-j)^2$.

A second test statistic can be formulated by noting that \eqref{b} implies that $X$ has a Pareto distribution if, and only if, 
\begin{equation}
 P(X>st)=P(X>s)P(X>t), \quad s,t\ge 1. \label{d}
\end{equation}
Again, estimating the left hand side of \eqref{d} nonparametrically by the empirical survival function and the right hand side by the parametric survival function with estimated shape parameter, we have the test statistic
\begin{align*}
MP^{(2)}_n &=\int_{1}^{\infty}\int_{1}^{\infty}\left(\frac{1}{n}\sum_{i=1}^{n}I(X_{j}>st)-(st)^{-\widehat{\beta}_n}\right)^2 \text{d}F_{\widehat{\beta}_n}(s) \text{d}F_{\widehat{\beta}_n}(t)\\
    &=\int_{1}^{\infty}\int_{1}^{\infty}\left(\frac{1}{n}\sum_{i=1}^{n}I(X_{j}>st)-(st)^{-\widehat{\beta}_n}\right)^2(st)^{-(\widehat{\beta}_n+1)} \widehat{\beta}_n^{\,2} \text{d}s\text{d}t.
\end{align*}
After some algebraic manipulation we have that
\begin{align*}
     MP^{(2)}_n=&\frac{10}{9}-\frac{1}{n^2}\sum_{j=1}^{n}\sum_{k=1}^{n}\min({X_j},{X_k})^{-\widehat{\beta}_n}-\frac{\widehat{\beta}_n}{n^2}\sum_{j=1}^{n}\sum_{k=1}^{n}\min({X_j},{X_k})^{-\widehat{\beta}_n}\log\bigl(\min({X_j},{X_k})\bigl)\\
     &-\frac{\widehat{\beta}_n}{n}\sum_{j=1}^{n}\left[\frac{1-{X_j}^{-2\widehat{\beta}_n}}{2\widehat{\beta}_n}-{X_j}^{-2\widehat{\beta}_n}\log({X_j})\right].
\end{align*}
Again this test statistic simplifies to a single sum involving order statistics
\begin{align*}
     MP^{(2)}_n=&\frac{10}{9}-\frac{1}{n^2}\sum_{j=1}^{n} v_{j,n} X_{(j)}^{-\widehat{\beta}_n}-\frac{\widehat{\beta}_n}{n^2}\sum_{j=1}^{n} v_{j,n} X_{(j)}^{-\widehat{\beta}_n}\log({X_{(j)}}) \\
     &-\frac{\widehat{\beta}_n}{n}\sum_{j=1}^{n}\left[\frac{1-{X_j}^{-2\widehat{\beta}_n}}{2\widehat{\beta}_n}-{X_j}^{-2\widehat{\beta}_n}\log({X_{(j)}})\right].
\end{align*}
Both $MP^{(1)}_n$ and $MP^{(2)}_n$ reject the hypothesis in \eqref{hyp} for large values.

\section{Monte Carlo study} \label{sect3}

In this section, we use Monte Carlo simulation to compare the performances of our newly proposed tests to some of the existing tests found in the statistical literature. 
The classical goodness-of-fit tests included are the Kolmogorov-Smirnov, Cram\'{e}r-von Mises and Anderson-Darling tests. 
Additionally, we include a test based on likelihood ratios as well as a test based on the Mellin transform. 
The calculable forms of the test statistics are specified below:
\begin{itemize}
   \item The Kolmogorov-Smirnov test:
   $$KS_n = \max\left\{\max_{1\leq j\leq n}\left[\frac{j}{n}-F_{\widehat{\beta}_n}(X_{(j)})\right], \max_{1\leq j\leq n}\left[F_{\widehat{\beta}_n}(X_{(j)})-\frac{j-1}{n}\right]\right\}.$$
   \item The Cram\'{e}r-von Mises test:
   $$ CM_n = \frac{1}{12n} +\sum_{j=1}^{n}\left[F_{\widehat{\beta}_n}(X_{(j)})-\frac{2j-1}{2n}\right]^2.$$
   \item The Anderson-Darling test:
   $$ AD_n = -n-\frac{1}{n}\sum_{j=1}^{n}(2j-1)\left[\log\left(F_{\widehat{\beta}_n}(X_{(j)})\right)+\log\left(1-F_{\widehat{\beta}_n}(X_{(n+1-j)})\right)\right].$$
   \item The test proposed by \cite{zhang2002powerful}, based on likelihood ratios: 
   $$ ZA_n=-\sum_{j=1}^{n}\left\{\frac{\log\left(F_{\widehat{\beta}_n}(X_{(j)})\right)}{n-j+\frac{1}{2}}+\frac{\log\left(1-F_{\widehat{\beta}_n}(X_{(j)})\right)}{j-\frac{1}{2}}\right\}.$$
   \item The test proposed by \cite{meintanis2009unified}, based on the Mellin transform:
\begin{eqnarray*}
    G_{n,a}&=&\frac{1}{n}\left[(\widehat{\beta}_n+1)^2\sum_{j,k=1}^{n}I^{(0)}_w({X_jX_k})+\sum_{j,k=1}^{n}I^{(2)}_w({X_jX_k})+2(\widehat{\beta}_n+1)\sum_{j,k=1}^{n}I^{(1)}_w({X_jX_k})\right]\\
    &&+\widehat{\beta}_n\left[n\widehat{\beta}_nI^{(0)}_w(1)-2(\widehat{\beta}_n+1)\sum_{j=1}^{n}I^{(0)}_w({X_j})-2\sum_{j=1}^{n}I^{(1)}_w({X_j})\right],
   \end{eqnarray*}
   where 
\begin{equation*}
    I^{(m)}_w(x)=\int_{0}^{\infty}(t-1)^m\frac{1}{x^t}w(t)\text{d}t, \quad m=0,1,2.
\end{equation*}
Choosing $w(t)=e^{-ax}$, one has
\begin{equation*}
    I^{(0)}_a(x)=(a+\log x)^{-1},
\end{equation*}
\begin{equation*}
    I^{(1)}_a(x)=\frac{1-a-\log x}{(a+\log x)^2},
\end{equation*}
and
\begin{equation*}
    I^{(2)}_a(x)=\frac{2-2a+a^2+2(a-1)\log x +\log^2 x}{(a+\log x)^3}.
\end{equation*}
To obtain the numerical results presented later in this section, the tuning parameter $a$ has been assigned a value of $1$.
\end{itemize}
All of the tests above reject the null hypothesis in (\ref{hyp}) for large values of the test statistics.
%
In addition to these tests for the Pareto distribution, we also compare the numerical powers achieved by the newly proposed tests against those of some tests for exponentiality (the interested reader is referred to \cite{allison2017apples} and \cite{henze2005recent} for an overview of tests for exponentiality). In these cases, the tests for exponentiality are applied to the log-transformed data 
(if $X \sim P(\beta)$ then $Y=\log X$ follows an exponential distribution with rate parameter $\beta$).

The exponentiality tests we consider are again the three classical tests based on the empirical distribution function (denoted by $\widetilde{KS}_n$, $\widetilde{CV}_n$, and $\widetilde{AD}_n$). 
In addition, we consider a test based on likelihood ratios by \cite{zhang2002powerful} ($\widetilde{ZA}_n$), 
a  test proposed by \cite{angus1982goodness} which is based on the memoryless property of the exponential distribution ($\widetilde{AG}_n$), 
a score based test proposed by \cite{cox1984analysis} ($\widetilde{CO}_n$), and 
a Cram\'{e}r-von Mises type test based on mean residuals life proposed by \cite{baringhaus2000tests} ($\widetilde{BH}_n$).

   
\subsection{Simulation setting}
We start by mentioning that the way we obtain the critical values differs depending on the estimation method.
%
When using the MLE in \eqref{betamle} to estimate $\beta$, the critical values of the tests considered do not depend on the value of $\beta$, i.e., if $X \sim P(\beta)$, then $Y := X^\beta \sim P(1)$. Consider the transformed sample $Y_j = X_j^{\widetilde{\beta}_n}, \ j=1,\dots,n$, and note that
\begin{equation}
    \widetilde{\beta}(Y_1,...,Y_n) = \frac{n}{\sum_{j=1}^{n}\log(Y_j)} = \frac{n}{\widetilde{\beta}(X_1,...,X_n)\sum_{j=1}^{n}\log(X_j)} = \frac{ \widetilde{\beta}(X_1,...,X_n)}{ \widetilde{\beta}(X_1,...,X_n)} = 1.
\end{equation}
For more details regarding this transformation and the estimation of critical values using Monte Carlo simulation, see Section 3.1 of \cite{ndwandwe2022testing}. We simulate $100\,000$ Monte Carlo samples (from a $P(1)$ distribution) to estimate these critical values. Next, we obtain $10\,000$ samples from an alternative distribution in order to obtain the empirical power against the distribution in question. Below, we calculate empirical powers against the distributions listed in Table \ref{tab1}. Note that the alternative distributions employed in the simulation study are right-shifted by one unit so as to ensure that they share the domain of the null distribution. 

In the case of MME, fixed critical values cannot be obtained because $\ddot{\beta}(Y_1,...,Y_n) \neq 1$. Instead we employ a parametric bootstrap approach to obtain the relevant empirical critical values. Due to the time-consuming nature of the bootstrap procedure, the warp-speed bootstrap technique proposed by \cite{giacomini2013warp} is used to obtain empirical powers. To this end, we generate $50\,000$ Monte Carlo samples and for each Monte Carlo sample we obtain one parametric bootstrap sample. The estimated empirical power is then the proportion of these $50\,000$ Monte Carlo samples that lead to the rejection of the null hypothesis in (\ref{hyp}), see \cite{ndwandwe2022testing} for an algorithm for the implementation of the warp-speed bootstrap in the current context. All calculations below are performed in R, see \cite{CRAN}.

\begin{table}[!htbp!]%
\begin{center}
\caption {Summary of various choices of the alternative distributions.}
\begin{tabular}{lll}
\hline
\hline
Alternative & Density function & Notation\\
\hline
Gamma & $\frac{1}{\Gamma(\theta)}(x-1)^{\theta-1}\mathrm{e}^{-(x-1)}$ & $\Gamma(\theta)$ \\
Weibull & $\theta (x-1)^{\theta-1}\exp(-(x-1)^\theta)$ & $W(\theta)$ \\
Log-normal & $\exp\left\{-\frac{1}{2}(\log(x-1)/\theta)^2\right\}/\left\{\theta (x-1) \sqrt{2\pi}\right\}$ & $LN(\theta)$ \\
Half-normal & $\frac{\sqrt{2}}{\theta\sqrt{\pi}}\exp\left(\frac{-(x-1)^2}{2\theta^2}\right)$ & $HN(\theta)$ \\
Linear failure rate & $(1+\theta (x-1))\exp(-(x-1)-\theta (x-1)^2/2)$ & $LFR(\theta)$ \\
Beta exponential & $\theta \mathrm{e}^{-(x-1)}(1-\mathrm{e}^{-(x-1)})^{\theta-1}$ &$BE(\theta)$\\
Tilted Pareto & $\frac{1+\theta}{(x+\theta)^2}$ & $TP (\theta)$\\
Dhillon & $\frac{\theta + 1}{x}\exp\left\{-(\log(x))^{\theta+1}\right\}(\log(x))^\theta$ & $D(\theta)$\\
\hline
\hline
\label{tab1}
\end{tabular}
\end{center}
\end{table}

\subsection{Powers against fixed alternatives}

For each of the tests considered, Tables \ref{tab2} and \ref{tab3} report power estimates for samples of size $n=20$ and $n=30$. These estimates are calculated against the alternative distributions given in Table \ref{tab1} for various parameter values. The resulting numerical powers are calculated to be the percentage, rounded to the nearest integer, of the samples that lead to the rejection of the null hypothesis.

Consider first the tests specifically developed for the Pareto distribution, i.e., those which do not require the data to be log-transformed. The tables include the results obtained using both the MLE and the MME. For ease of comparison, we circle the highest two 
numerical powers associated with the MME and we indicate the two highest powers achieved using the MLE with a rectangle (including ties). 
In each case, the best performing test overall is indicated using shading as well. 
Additionally, 
these two 
power tables include results pertaining to the tests based on the transformed data. In the majority of the cases considered, the $\widetilde{CO}_n$ test performs best among the tests for exponentiality. When comparing the powers of the tests developed for the Pareto distribution to those based on the transformed data, the powers of the former group are found to be substantially higher. As a result, we do not report the powers achieved by all of the tests based on log-transformed data in Tables \ref{tab2} and \ref{tab3}, we only indicate the best performing test from this group together with the associated power; this is shown in the last column of each table. In cases where the tests based on the exponential distribution outperform or equal the performance of the others, these powers are indicated using shading. It should be noted that outright outperformance of this kind is only achieved against two alternative distributions; the $LN(2.5)$ (the $\widetilde{CO}_n$ performs best in this case) and the $HN(0.5)$ (where the $\widetilde{BH}_n$ test proves most powerful). In both of the mentioned cases, the outperformance is by a narrow margin. For completeness, we include the numerical powers associated with the exponentiality tests in the appendix. However, these results are not discussed further, except to point out that the newly proposed tests based on the multiplicative memoryless property substantially outperform the test based on the memoryless property of the exponential distribution against the considered alternatives.

\begin{table}[htbp]
  \centering
 \caption{Numerical powers against fixed alternatives with $n=20$}
  \resizebox{\textwidth}{!}{
    \begin{tabular}{|l|rr|rr|rr|rr|rr|rr|rr||r|}
    \cline{1-16}          	&	 \multicolumn{2}{c|}{$KS_n$} 	&	 \multicolumn{2}{c|}{$CV_n$} 	&	 \multicolumn{2}{c|}{$AD_n$} 	&	 \multicolumn{2}{c|}{$ZA_n$} 	&	 \multicolumn{2}{c|}{$G_{n,1}$} 	&	 \multicolumn{2}{c|}{$MP^{(1)}_n$} 	&	 \multicolumn{2}{c||}{$MP^{(2)}_n$} &	 \multicolumn{1}{c|}{Exp.} \bigstrut\\	
    \cline{1-15}
     Distribution     	&	 \multicolumn{1}{c}{MME} 	&	 \multicolumn{1}{c|}{MLE} 	&	 \multicolumn{1}{c}{MME} 	&	 \multicolumn{1}{c|}{MLE} 	&	 \multicolumn{1}{c}{MME} 	&	 \multicolumn{1}{c|}{MLE} 	&	 \multicolumn{1}{c}{MME} 	&	 \multicolumn{1}{c|}{MLE} 	&	 \multicolumn{1}{c}{MME} 	&	 \multicolumn{1}{c|}{MLE}  	&	 \multicolumn{1}{c}{MME} 	&	 \multicolumn{1}{c|}{MLE} 	&	 \multicolumn{1}{c}{MME} 	&	 \multicolumn{1}{c||}{MLE}   &   \multicolumn{1}{c|}{test} 	\\
     \hline																													
    $P(2)$  	&	 \textbf{5} 	&	 \textbf{5} 	&	 \textbf{5} 	&	 \textbf{5} 	&	 \textbf{5} 	&	 \textbf{5} 	&	 \textbf{5} 	&	 \textbf{5} 	&	 \textbf{5} 	&	 \textbf{5}  	&	 \textbf{5} 	&	 \textbf{5} 	&	 \textbf{5} 	&	 \textbf{5} 	&	 \textbf{5} ($all$)\\
    $P(5)$  	&	 \textbf{5} 	&	 \textbf{5} 	&	 \textbf{5} 	&	 \textbf{5} 	&	 \textbf{5} 	&	 \textbf{5} 	&	 \textbf{5} 	&	 \textbf{5} 	&	 \textbf{5} 	&	 \textbf{5}  	&	 \textbf{5} 	&	 \textbf{5} 	&	 \textbf{5} 	&	 \textbf{5} 	&	 \textbf{5} ($all$)\\
    $P(10)$ 	&	 \textbf{5} 	&	 \textbf{5} 	&	 \textbf{5} 	&	 \textbf{5} 	&	 \textbf{5} 	&	 \textbf{5} 	&	 \textbf{5} 	&	 \textbf{5} 	&	4	&	4	&	 \textbf{5} 	&	 \textbf{5} 	&	 \textbf{5} 	&	 \textbf{5} 	
    & \textbf{5} ($all$) \\
    $\Gamma(0.8)$ 	&	15	&	9	&	16	&	10	&	15	&	10	&	11	&	11	&	 \textbf{\circled{19}} 	&	10	&	18	&	 \textbf{\squared{12}} 	&	 \textbf{\circledfill{21}}   	&	\textbf{\squared{14}} 	
    & \textbf{13} ($\widetilde{CO}_n$)\\
    $\Gamma(1)$ 	&	37	&	24	&	44	&	29	&	42	&	24	&	33	&	28	&	 \textbf{\circledfill{51}} 	&	 \textbf{\squared{32}}  	&	47	&	31	&	 \textbf{\circled{48}} 	&	 \textbf{\squared{32}} 
    & \textbf{38} ($\widetilde{CO}_n$)\\
    $\Gamma(1.2)$ 	&	64	&	45	&	74	&	56	&	72	&	51	&	64	&	55	&	     \textbf{\circledfill{80}} 	&	     \textbf{\squared{60}}  	&	 \textbf{\circled{75}} 	&	     \textbf{\squared{57}} 	&	     \textbf{\circled{75}} 	&	     \textbf{\squared{57}} 	
    & \textbf{69} ($\widetilde{CO}_n$)\\
    $W(0.8)$ 	&	15	&	9	&	17	&	9	&	17	&	9	&	11	&	 \textbf{\squared{10}} 	&	 \textbf{\circled{20}} 	&	8	&	19	&	 \textbf{\squared{10}} 	&	 \textbf{\circledfill{22}} 	&	 \textbf{\squared{12}} 	
    & \textbf{9} ($\widetilde{CO}_n$)\\
    $W(1.2)$ 	&	65	&	50	&	75	&	62	&	73	&	57	&	65	&	60	&	 \textbf{\circledfill{80}} 	&	 \textbf{\squared{65}}  	&	  \textbf{\circled{76}} 	&	 \textbf{\squared{63}} 	&	\textbf{\circled{76}} 	&	62	
    & \textbf{73} ($\widetilde{CO}_n$)\\
    $W(1.5)$ 	&	91	&	82	&	96	&	92	&	96	&	90	&	93	&	91	&	 \textbf{\circledfill{98}} 	&	 \textbf{\squared{94}}  	&	  \textbf{\circled{97}} 	&	 \textbf{\squared{93}} 	&	96	&	92	
    & \textbf{96} ($\widetilde{CO}_n$)\\
    $LN(1)$  	&	72	&	55	&	82	&	66	&	83	&	64	&	 \textbf{\circledfill{90}} 	&	 \textbf{\squared{80}} 	&	 \textbf{\circled{85}} 	&	 \textbf{\squared{72}}  	&	80	&	62	&	72	&	51	
    & \textbf{88} ($\widetilde{CO}_n$)\\
    $LN(1.5)$ 	&	17	&	6	&	20	&	7	&	22	&	6	&	21	&	 \textbf{\squared{10}} 	&	 \textbf{\circledfill{24}} 	&	 \textbf{\squared{8}}  	&	22	&	7	&	 \textbf{\circled{23}} 	&	7	
    & \textbf{14} ($\widetilde{CO}_n$)\\
    $LN(2.5)$ 	&	11	&	27	&	9	&	29	&	 \textbf{\circled{26}} 	&	 \textbf{\squaredfill{43}} 	&	 \textbf{\circled{22}} 	&	27	&	5	&	 \textbf{\squared{32}}  	&	14	&	26	&	17	&	17	
    &  \cellcolor{black!30}\textbf{47} ($\widetilde{CO}_n$)\\
    $HN(0.5)$ 	&	44	&	37	&	53	&	47	&	49	&	41	&	43	&	43	&	55	&	47	&	 \textbf{\circled{56}} 	&	 \textbf{\squared{50}} 	&	 \textbf{\circledfill{57}} 	&	 \textbf{\squared{51}} 	
    &  \cellcolor{black!30}\textbf{63} ($\widetilde{BH}_n$)\\
    $HN(1)$ 	&	67	&	54	&	76	&	65	&	74	&	59	&	62	&	60	&	 \textbf{\circledfill{80}} 	&	66	&	78	&	 \textbf{\squared{68}} 	&	 \textbf{\circledfill{80}} 	&	 \textbf{\squared{70}} 	
    & \textbf{69} ($\widetilde{BH}_n$)\\
    $HN(1.2)$ 	&	74	&	58	&	82	&	70	&	80	&	64	&	69	&	66	&	 \textbf{\circledfill{86}} 	&	71	&	84	&	 \textbf{\squared{73}} 	&	 \textbf{\circledfill{86}} 	&	 \textbf{\squared{76}} 	
    & \textbf{76} ($\widetilde{BH}_n$)\\
    $LFR(0.2)$ 	&	45	&	31	&	52	&	38	&	49	&	33	&	40	&	35	&	 \textbf{\circledfill{59}} 	&	 \textbf{\squared{41}}  	&	55	&	 \textbf{\squared{41}} 	&	 \textbf{\circled{57}} 	&	 \textbf{\squared{43}} 	
    & \textbf{47} ($\widetilde{CO}_n$)\\
    $LFR(0.8)$ 	&	54	&	42	&	64	&	53	&	60	&	47	&	51	&	49	&	 \textbf{\circledfill{69}} 	&	55	&	66	&	 \textbf{\squared{56}} 	&	 \textbf{\circled{68}} 	&	 \textbf{\squared{57}} 	
    & \textbf{58} ($\widetilde{CO}_n$)\\
    $LFR(1)$ 	&	56	&	45	&	65	&	55	&	62	&	50	&	53	&	52	&	 \textbf{\circledfill{70}} 	&	57	&	68	&	 \textbf{\squared{58}} 	&	 \textbf{\circled{69}} 	&	 \textbf{\squared{60}} 	
    & \textbf{61} ($\widetilde{CO}_n$)\\
    $BE(0.8)$ 	&	17	&	11	&	19	&	12	&	18	&	11	&	13	&	12	&	 \textbf{\circled{23}} 	&	12	&	22	&	 \textbf{\squared{13}} 	&	 \textbf{\circledfill{25}} 	&	 \textbf{\squared{16}} 	
    & \textbf{13} ($\widetilde{CO}_n$)\\
    $BE(1)$ 	&	39	&	25	&	46	&	31	&	43	&	25	&	34	&	28	&	 \textbf{\circledfill{52}} 	&	 \textbf{\squared{33}}  	&	48	&	32	&	 \textbf{\circled{50}} 	&	 \textbf{\squared{34}} 	
    & \textbf{39} ($\widetilde{CO}_n$)\\
    $BE(1.5)$ 	&	84	&	67	&	 \textbf{\circled{91}} 	&	 \textbf{\squared{80}} 	&	 \textbf{\circled{91}} 	&	75	&	86	&	 \textbf{\squared{80}} 	&	 \textbf{\circledfill{94}} 	&	 \textbf{\squared{83}}  	&	 \textbf{\circled{91}} 	&	 \textbf{\squared{80}} 	&	90	&	77	
    & \textbf{89} ($\widetilde{CO}_n$)\\
    $TP(1)$ 	&	46	&	12	&	53	&	 \textbf{\squared{13}} 	&	 \textbf{\circledfill{58}} 	&	10	&	51	&	 \textbf{\squared{13}} 	&	54	&	 \textbf{\squared{13}}  	&	57	&	 \textbf{\squared{13}} 	&	 \textbf{\circledfill{58}} 	&	 \textbf{\squared{13}} 	
    & \textbf{19} ($\widetilde{CO}_n$)\\
    $TP(2)$ 	&	76	&	22	&	82	&	 \textbf{\squared{26}} 	&	 \textbf{\circledfill{86}} 	&	21	&	82	&	25	&	82	&	25	&	85	&	 \textbf{\squared{26}} 	&	 \textbf{\circledfill{86}} 	&	24	
    & \textbf{34} ($\widetilde{CO}_n$)\\
    $TP(3)$ 	&	90	&	32	&	94	&	 \textbf{\squared{37}} 	&	 \textbf{\circledfill{96}} 	&	32	&	94	&	35	&	94	&	36	&	 \textbf{\circledfill{96}} 	&	 \textbf{\squared{38}} 	&	 \textbf{\circledfill{96}} 	&	36	
    & \textbf{47} ($\widetilde{CO}_n$)\\
    $D(0.4)$ 	&	50	&	29	&	 \textbf{\circled{60}} 	&	34	&	 \textbf{\circled{60}} 	&	30	&	57	&	 \textbf{\squared{39}} 	&	 \textbf{\circledfill{64}} 	&	 \textbf{38}  	&	59	&	33	&	57	&	30	
    & \textbf{52} ($\widetilde{CO}_n$)\\
    $D(0.6)$ 	&	73	&	51	&	82	&	61	&	 \textbf{\circled{83}} 	&	56	&	80	&	 \textbf{\squared{65}} 	&	 \textbf{\circledfill{86}} 	&	 \textbf{\squared{67}}  	&	82	&	59	&	79	&	54	
    & \textbf{80} ($\widetilde{CO}_n$)\\
    $D(0.8)$ 	&	89	&	70	&	 \textbf{\circled{95}} 	&	81	&	94	&	79	&	93	&	 \textbf{\squared{84}} 	&	 \textbf{\circledfill{96}} 	&	 \textbf{\squared{87}}  	&	94	&	81	&	92	&	76	
    & \textbf{93} ($\widetilde{CO}_n$)\\
\hline   \end{tabular}%
}
\label{tab2} 
\end{table}%

\begin{table}[htbp]
  \centering
 \caption{Numerical powers against fixed alternatives with $n=30$}
 \resizebox{\textwidth}{!}{
    \begin{tabular}{|l|rr|rr|rr|rr|rr|rr|rr||r|}
    \cline{1-16}          	&	 \multicolumn{2}{c|}{$KS_n$} 	&	 \multicolumn{2}{c|}{$CV_n$} 	&	 \multicolumn{2}{c|}{$AD_n$} 	&	 \multicolumn{2}{c|}{$ZA_n$} 	&	 \multicolumn{2}{c|}{$G_{n,1}$} 	&	 \multicolumn{2}{c|}{$MP^{(1)}_n$} 	&	 \multicolumn{2}{c||}{$MP^{(2)}_n$} &	 \multicolumn{1}{c|}{Exp.} \bigstrut\\	
    \cline{1-15}
     Distribution     	&	 \multicolumn{1}{c}{MME} 	&	 \multicolumn{1}{c|}{MLE} 	&	 \multicolumn{1}{c}{MME} 	&	 \multicolumn{1}{c|}{MLE} 	&	 \multicolumn{1}{c}{MME} 	&	 \multicolumn{1}{c|}{MLE} 	&	 \multicolumn{1}{c}{MME} 	&	 \multicolumn{1}{c|}{MLE} 	&	 \multicolumn{1}{c}{MME} 	&	 \multicolumn{1}{c|}{MLE}  	&	 \multicolumn{1}{c}{MME} 	&	 \multicolumn{1}{c|}{MLE} 	&	 \multicolumn{1}{c}{MME} 	&	 \multicolumn{1}{c||}{MLE}   &   \multicolumn{1}{c|}{test} 	\\
     \hline	
    $P(2)$  	&	 \textbf{5} 	&	 \textbf{5} 	&	 \textbf{5} 	&	 \textbf{5} 	&	 \textbf{5} 	&	 \textbf{5} 	&	 \textbf{5} 	&	 \textbf{5} 	&	 \textbf{5} 	&	 \textbf{5}  	&	 \textbf{5} 	&	 \textbf{5} 	&	 \textbf{5} 	&	 \textbf{5} 	&	 \textbf{5} ($all$)\\
    $P(5)$  	&	 \textbf{5} 	&	 \textbf{5} 	&	 \textbf{5} 	&	 \textbf{5} 	&	 \textbf{5} 	&	 \textbf{5} 	&	 \textbf{5} 	&	 \textbf{5} 	&	 \textbf{5} 	&	 \textbf{5}  	&	 \textbf{5} 	&	 \textbf{5} 	&	 \textbf{5} 	&	 \textbf{5} 	&	 \textbf{5} ($all$)\\
    $P(10)$ 	&	 \textbf{5} 	&	 \textbf{5} 	&	 \textbf{5} 	&	 \textbf{5} 	&	 \textbf{5} 	&	 \textbf{5} 	&	 \textbf{5} 	&	 \textbf{5} 	&	 \textbf{5} 	&	 \textbf{5}  	&	 \textbf{5} 	&	 \textbf{5} 	&	 \textbf{5} 	&	 \textbf{5} 	& \textbf{5} ($all$)\\
    $\Gamma(0.8)$ 	&	19	&	12	&	21	&	14	&	21	&	13	&	14	&	14	&	 \textbf{\circled{25}} 	&	14	&	 \textbf{\circled{25}} 	&	 \textbf{\squared{16}} 	&	 \textbf{\circledfill{29}} 	&	 \textbf{\squared{19}} 	& {14} ($\widetilde{CO}_n$)\\
    $\Gamma(1)$ 	&	53	&	36	&	61	&	44	&	59	&	39	&	47	&	42	&	 \textbf{\circledfill{67}} 	&	 \textbf{\squared{49}}  	&	64	&	47	&	 \textbf{\circled{65}} 	&	 \textbf{\squared{48}} 	& {53} ($\widetilde{CO}_n$)\\
    $\Gamma(1.2)$ 	&	82	&	64	&	89	&	76	&	89	&	73	&	82	&	76	&	 \textbf{\circledfill{93}} 	&	 \textbf{\squared{80}}  	&	 \textbf{\circled{90}} 	&	 \textbf{\squared{78}} 	&	 \textbf{\circled{90}} 	&	77	& {85} ($\widetilde{CO}_n$)\\
    $W(0.8)$ 	&	20	&	10	&	21	&	11	&	21	&	11	&	13	&	12	&	23	&	10	&	 \textbf{\circled{25}} 	&	 \textbf{\squared{13}} 	&	 \textbf{\circledfill{29}} 	&	 \textbf{\squared{16}} 	& {11} ($\widetilde{CO}_n$)\\
   $ W(1.2)$ 	&	83	&	69	&	90	&	81	&	90	&	78	&	83	&	81	&	 \textbf{\circledfill{93}} 	&	 \textbf{\squared{85}}  	&	 \textbf{\circled{91}} 	&	 \textbf{\squared{83}} 	&	90	&	81	& {89} ($\widetilde{CO}_n$)\\
    $W(1.5)$ 	&	99	&	95	&	 \textbf{\circledfill{100}} 	&	 \textbf{\squared{99}} 	&	 \textbf{\circledfill{100}} 	&	 \textbf{\squared{99}} 	&	99	&	 \textbf{\squared{99}} 	&	 \textbf{\circledfill{100}} 	&	 \textbf{\squared{99}}  	&	 \textbf{\circledfill{100}} 	&	 \textbf{\squared{99}} 	&	 \textbf{\circledfill{100}} 	&	 \textbf{\squared{99}} 	& \cellcolor{black!30}\textbf{100} ($\widetilde{CO}_n$)\\
    $LN(1)$  	&	88	&	75	&	94	&	85	&	 \textbf{\circled{95}} 	&	87	&	 \textbf{\circledfill{98}} 	&	 \textbf{\squared{95}} 	&	 \textbf{\circled{95}} 	&	 \textbf{\squared{90}}  	&	92	&	82	&	87	&	71	& \cellcolor{black!30}\textbf{98} ($\widetilde{CO}_n$)\\
    $LN(1.5)$ 	&	21	&	8	&	26	&	8	&	 \textbf{\circled{28}} 	&	7	&	 \textbf{\circled{28}} 	&	 \textbf{\squared{14}} 	&	 \textbf{\circledfill{30}} 	&	 \textbf{\squared{10}}  	&	 \textbf{\circled{28}} 	&	8	&	 \textbf{\circled{28}} 	&	8	& {20} ($\widetilde{CO}_n$)\\
    $LN(2.5)$ 	&	13	&	38	&	10	&	42	&	 \textbf{\circled{31}} 	&	 \textbf{\squaredfill{58}} 	&	 \textbf{\circled{30}} 	&	39	&	4	&	 \textbf{\squared{44}}  	&	16	&	39	&	19	&	26	&  \cellcolor{black!30}\textbf{61} ($\widetilde{CO}_n$)\\
    $HN(0.5)$ 	&	61	&	54	&	71	&	66	&	68	&	61	&	61	&	63	&	73	&	68	&	 \textbf{\circled{74}} 	&	 \textbf{\squared{69}} 	&	 \textbf{\circledfill{75}} 	&	 \textbf{\squared{70}} 	&  \cellcolor{black!30}\textbf{80} ($\widetilde{BH}_n$)\\
    $HN(1)$ 	&	85	&	73	&	91	&	84	&	90	&	81	&	82	&	83	&	 \textbf{\circledfill{93}} 	&	86	&	92	&	 \textbf{\squared{87}} 	&	 \textbf{\circledfill{93}} 	&	 \textbf{\squared{88}} 	& {84} ($\widetilde{BH}_n$)\\
    $HN(1.2)$ 	&	90	&	78	&	95	&	88	&	94	&	86	&	87	&	88	&	 \textbf{\circledfill{96}} 	&	90	&	 \textbf{\circledfill{96}} 	&	 \textbf{\squared{91}} 	&	\textbf{\circledfill{96}} 	&	 \textbf{\squared{92}} 	& {90} ($\widetilde{BH}_n$)\\
    $LFR(0.2)$ 	&	63	&	47	&	71	&	57	&	69	&	52	&	56	&	54	&	 \textbf{\circledfill{77}} 	&	 \textbf{\squared{61}}  	&	74	&	60	&	 \textbf{\circled{75}} 	&	 \textbf{\squared{62}} 	& {62} ($\widetilde{CO}_n$)\\
    $LFR(0.8)$ 	&	73	&	60	&	82	&	73	&	80	&	69	&	70	&	71	&	 \textbf{\circledfill{85}} 	&	 \textbf{\squared{76}}  	&	84	&	 \textbf{\squared{76}} 	&	 \textbf{\circledfill{85}} 	&	 \textbf{\squared{78}} 	& {76} ($\widetilde{CO}_n$)\\
    $LFR(1)$ 	&	74	&	63	&	83	&	75	&	81	&	71	&	72	&	73	&	 \textbf{\circledfill{86}} 	&	 \textbf{\squared{78}}  	&	85	&	 \textbf{\squared{78}} 	&	\textbf{\circledfill{86}} 	&	 \textbf{\squared{79}} 	& {79} ($\widetilde{CO}_n$)\\
    $BE(0.8)$ 	&	24	&	15	&	27	&	17	&	26	&	15	&	16	&	17	&	 \textbf{\circled{31}} 	&	17	&	30	&	 \textbf{\squared{19}} 	&	 \textbf{\circledfill{34}} 	&	 \textbf{\squared{23}} 	& {17} ($\widetilde{CO}_n$)\\
    $BE(1)$ 	&	52	&	36	&	60	&	44	&	59	&	39	&	46	&	42	&	 \textbf{\circledfill{67}} 	&	 \textbf{\squared{48}}  	&	63	&	46	&	 \textbf{\circled{64}} 	&	 \textbf{\squared{48}} 	& {52} ($\widetilde{CO}_n$)\\
    $BE(1.5)$ 	&	95	&	86	&	98	&	94	&	 \textbf{\circledfill{99}} 	&	93	&	97	&	 \textbf{\squared{95}} 	&	 \textbf{\circledfill{99}} 	&	 \textbf{\squared{96}}  	&	98	&	 \textbf{\squared{95}} 	&	98	&	93	& {98} ($\widetilde{CO}_n$)\\
    $TP(1)$ 	&	58	&	15	&	66	&	 \textbf{\squared{17}} 	&	 \textbf{\circledfill{70}} 	&	14	&	64	&	 \textbf{\squared{17}} 	&	65	&	 \textbf{\squared{17}}  	&	69	&	 \textbf{\squared{17}} 	&	 \textbf{\circledfill{70}} 	&	16	& {25} ($\widetilde{CO}_n$)\\
    $TP(2)$ 	&	88	&	30	&	93	&	 \textbf{\squared{37}} 	&	 \textbf{\circledfill{95}} 	&	32	&	93	&	33	&	92	&	 \textbf{\squared{37}}  	&	 \textbf{\circled{94}} 	&	 \textbf{\squared{37}} 	&	 \textbf{\circled{94}} 	&	34	& {47} ($\widetilde{CO}_n$)\\
    $TP(3)$ 	&	97	&	45	&	 \textbf{\circledfill{99}} 	&	 \textbf{\squared{54}} 	&	 \textbf{\circledfill{99}} 	&	49	&	 \textbf{\circledfill{99}} 	&	47	&	 \textbf{\circledfill{99}} 	&	 \textbf{\squared{53}}  	&	 \textbf{\circledfill{99}} 	&	 \textbf{\squared{53}} 	&	 \textbf{\circledfill{99}} 	&	51	& {63} ($\widetilde{CO}_n$)\\
    $D(0.4)$ 	&	66	&	41	&	75	&	51	&	 \textbf{\circled{76}} 	&	48	&	73	&	 \textbf{\squared{56}} 	&	 \textbf{\circledfill{78}} 	&	 \textbf{\squared{57}}  	&	74	&	49	&	71	&	43	& {70} ($\widetilde{CO}_n$)\\
    $D(0.6)$ 	&	89	&	70	&	 \textbf{\squared{94}} 	&	81	&	 \textbf{\circled{94}} 	&	80	&	93	&	 \textbf{\squared{84}} 	&	 \textbf{\circledfill{96}} 	&	 \textbf{\squared{86}}  	&	 \textbf{\circled{94}} 	&	80	&	92	&	73	& {94} ($\widetilde{CO}_n$)\\
    $D(0.8)$ 	&	97	&	89	&	 \textbf{\circledfill{99}} 	&	95	&	 \textbf{\circledfill{99}} 	&	95	&	 \textbf{\circledfill{99}} 	&	 \textbf{\squared{96}} 	&	 \textbf{\circledfill{99}} 	&	 \textbf{\squared{97}}  	&	 \textbf{\circledfill{99}} 	&	95	&	98	&	92	& \cellcolor{black!30}\textbf{99} ($\widetilde{CO}_n$)\\

\hline    \end{tabular}%
}
\label{tab3}
\end{table}%

Tables \ref{tab2} and \ref{tab3} indicate that all of the tests considered achieve the specified size of $5\%$ against each of the Pareto distributions included in the Monte Carlo study. Unsurprisingly, the powers reported in Table \ref{tab3} are generally higher than those reported in Table \ref{tab2}, indicating that an increase in sample size results in an increase in power against the alternatives considered.

Consider the result obtained using the MLE. The tables indicate that, using this estimation technique, $MP_n^{(1)}$ and $G_{n,1}$ perform best, closely followed by $MP_n^{(2)}$. The $ZA_n$ and $CV_n$ tests also generally perform well in this setting. When turning our attention to the results obtained using the MME, we see that $MP^{(2)}_n$ and $G_{n,1}$ exhibit the highest powers, followed by $MP^{(1)}_n$ and $AD_n$.

We now compare the results obtained using the MLE to those based on the MME. 
Although the power achieved by a given test using the MLE is higher than that achieved using the MME in a few cases, the reverse is generally true. 
In the vast majority of cases, the results obtained using the MME is superior to those associated with the MLE. 
In some cases this outperformance is substantial. 
For example, the observed power of the $AD_n$ against the $TP(2)$ distribution is $86\%$ when using the MME, compared to a mere $21\%$ using the MLE.
Taking all of the observations made above into consideration, our recommendation when testing for the Pareto distribution 
is to use either the $MP_n^{(2)}$ or $G_{n,1}$ tests 
together with the MME.


\subsection{Powers against mixture distributions}

In addition to the powers obtained against fixed alternatives, we include results relating to two families of mixture distributions. Consider a mixture distribution with mixing proportion $p$, contaminated by some distribution $G$. When simulating from this distribution, we generate a realisation from $G$ with probability $p\in[0,1]$ and we generate a Pareto distributed random variate with probability $1-p$. The shape parameter of the Pareto distribution used is chosen so that the means of the two distributions under consideration are equal. The two contaminating distributions considered are the shifted exponential and half-normal distributions, each with mean $3$.
The calculated numerical powers can be found in Tables \ref{tab4} and \ref{tab5}. Again, to ease our comparison, we encircle the highest numerical power linked to the MME and mark the highest power attained through the MLE with a rectangle, including instances of ties. Additionally, we shade the highest power overall. Note that, in Tables \ref{tab4} and \ref{tab5}, the results associated with sample sizes $n=20$ (top) and $n=30$ (bottom) are included in the same table.

\begin{table}[htbp]
  \centering
 \caption{Numerical powers against exponential mixture distributions}
  \resizebox{\textwidth}{!}{
    \begin{tabular}{|c|r|rr|rr|rr|rr|rr|rr|rr|}
    \hline
          & \multicolumn{1}{l|}{$n$} & \multicolumn{2}{c|}{$KS_n$} & \multicolumn{2}{c|}{$CV_n$} & \multicolumn{2}{c|}{$AD_n$} & \multicolumn{2}{c|}{$ZA_n$} & \multicolumn{2}{c|}{$G_{n,1}$} & \multicolumn{2}{c|}{$MP^{(1)}_n$} & \multicolumn{2}{c|}{$MP^{(2)}_n$} \bigstrut\\
    \multicolumn{1}{|l|}{Proportion} &       & \multicolumn{1}{c}{MME} & \multicolumn{1}{c|}{MLE} & \multicolumn{1}{c}{MME} & \multicolumn{1}{c|}{MLE} & \multicolumn{1}{c}{MME} & \multicolumn{1}{c|}{MLE} & \multicolumn{1}{c}{MME} & \multicolumn{1}{c|}{MLE} & \multicolumn{1}{c}{MME} & \multicolumn{1}{c|}{MLE} & \multicolumn{1}{c}{MME} & \multicolumn{1}{c|}{MLE} & \multicolumn{1}{c}{MME} & \multicolumn{1}{c|}{MLE} \bigstrut\\
    \hline
    \multirow{2}[2]{*}{0.1} & 20    & \textbf{\circledfill{6}} & \textbf{\squared{5}} & \textbf{\circledfill{6}} & \textbf{\squared{5}} & 5     & \textbf{\squared{5}} & 5     & \textbf{\squared{5}} & 5 & \textbf{\squared{5}} & \textbf{\circledfill{6}} & \textbf{\squared{5}}  & \textbf{\circledfill{6}} & \textbf{\squared{5}}  \bigstrut[t]\\

          & 30    & 5 & 5 &5 & 5 & 5 & 5 & \textbf{\circledfill{6}} & \textbf{\squaredfill{6}} & 5 & 5 & 5 & 5 & 
           5 & 5  \bigstrut[b]\\
\hline

    \multirow{2}[2]{*}{0.3} & 20    & 6     & \textbf{\squared{6}} & 6     & \textbf{\squared{6}} & 6     & 5     & \textbf{\circledfill{7}} & \textbf{\squared{6}} & \textbf{\circledfill{7}} & 5 & \textbf{\circledfill{7}} & \textbf{\squared{6}}     & \textbf{\circledfill{7}} & \textbf{\squared{6}} 
     \bigstrut[t]\\
          & 30    & 6     & 6     & \textbf{\circledfill{7}} & 6     & 6     & 6     & \textbf{\circledfill{7}} & \textbf{\squaredfill{7}} & \textbf{\circledfill{7}} & 6 &\textbf{\circledfill{7}} & 6     & \textbf{\circledfill{7}} & \textbf{\squaredfill{7}}  \bigstrut[b]\\
 
         \hline
  
    \multirow{2}[2]{*}{0.5} & 20    & 8     & 6     & 8     & 7 & 7     & 6     & 8     & 7 & \textbf{\circledfill{9}} & 6 & 8     & 7     & \textbf{\circledfill{9}} & \textbf{\squared{8}}      \bigstrut[t]\\
    
          & 30    & 9     & 8     & 10 & \textbf{\squared{9}} & 9     & 7     & 9     & \textbf{\squared{9}} & 10 & 8 & 10 & \textbf{\squared{9}} & \textbf{\circledfill{11}} & \textbf{\squared{9}} \bigstrut[b]\\
   
      \hline
    
    \multirow{2}[2]{*}{0.7} & 20    & 10    & 9     & 12    & \textbf{\squared{10}} & 10    & 8     & 11    & \textbf{\squared{10}} & \textbf{\circledfill{13}} & 9& 12    & \textbf{\squared{10}}    & \textbf{\circledfill{13}} & \textbf{\squared{10}}     \bigstrut[t]\\
          & 30    & 13    & 11    & 15    & 12    & 13    & 11    & 13    & 13& 16 & 13 & 16 & 13 & \textbf{\circledfill{17}} & \textbf{\squared{14}}  \bigstrut[b]\\
   
         \hline

    \multirow{2}[2]{*}{0.9} & 20    & 14    & 11    & 16    & 13 & 14    & 10    & 13    & 13 & \textbf{\circledfill{18}} & 13& 17 & 13    & 17 & \textbf{\squared{14}}     \bigstrut[t]\\
          & 30    & 18    & 14    & 21    & 17    & 19    & 14    & 17    & 17    & \textbf{\circledfill{23}} & 18& 22    & 18 & \textbf{\circledfill{23}} & \textbf{\squared{19}}  \bigstrut[b]\\
\hline    \end{tabular}%
}
\label{tab4}
\end{table}

\begin{table}[htbp]
  \centering
 \caption{Numerical powers against lognormal mixture distributions}
  \resizebox{\textwidth}{!}{
    \begin{tabular}{|c|r|rr|rr|rr|rr|rr|rr|rr|}
    \hline
          & \multicolumn{1}{l|}{$n$} & \multicolumn{2}{c|}{$KS_n$} & \multicolumn{2}{c|}{$CV_n$} & \multicolumn{2}{c|}{$AD_n$} & \multicolumn{2}{c|}{$ZA_n$} & \multicolumn{2}{c|}{$G_{n,1}$} & \multicolumn{2}{c|}{$MP^{(1)}_n$} & \multicolumn{2}{c|}{$MP^{(2)}_n$} \bigstrut\\   \multicolumn{1}{|l|}{Proportion} &       & \multicolumn{1}{c}{MME} & \multicolumn{1}{c|}{MLE} & \multicolumn{1}{c}{MME} & \multicolumn{1}{c|}{MLE} & \multicolumn{1}{c}{MME} & \multicolumn{1}{c|}{MLE} & \multicolumn{1}{c}{MME} & \multicolumn{1}{c|}{MLE} & \multicolumn{1}{c}{MME} & \multicolumn{1}{c|}{MLE} & \multicolumn{1}{c}{MME} & \multicolumn{1}{c|}{MLE} & \multicolumn{1}{c}{MME} & \multicolumn{1}{c|}{MLE} \bigstrut\\
    \hline
    \multirow{2}[0]{*}{0.1} & 20    & \textbf{\circledfill{6}} & \textbf{\squaredfill{6}} & \textbf{\circledfill{6}} & 5     & 5     & 5     & \textbf{\circledfill{6}} & \textbf{\squaredfill{6}} & \textbf{\circledfill{6}} & 5 & \textbf{\circledfill{6}} & \textbf{\squaredfill{6}} & \textbf{\circledfill{6}} & \textbf{\squaredfill{6}}  \\
          & 30    & 5     & 5     & \textbf{\circledfill{6}} & 5     & 5     & 5     & \textbf{\circledfill{6}} & \textbf{\squaredfill{6}} & \textbf{\circledfill{6}} & 5 & \textbf{\circledfill{6}} & \textbf{\squaredfill{6}} & \textbf{\circledfill{6}} & \textbf{\squaredfill{6}}  \\
\hline
    \multirow{2}[0]{*}{0.3} & 20    & 10    & 8     & 10    & 9 & 9     & 7     & 9     & 9 & 11 & 8 & 11 & 9 & \textbf{\circledfill{12}} & \textbf{\squared{10}}  \\
          & 30    & 11    & 10    & 13    & 11 & 11    & 9     & 11    & 11 & 13    & 11 & 13    & 11 & \textbf{\circledfill{14}} & \textbf{\squared{13}}  \\

          \hline
    \multirow{2}[0]{*}{0.5} & 20    & 16    & 13    & 19    & 15    & 16    & 12    & 15    & 15    & 20 & 15 & 20 & 17 & \textbf{\circledfill{21}} & \textbf{\squared{18}}  \\
          & 30    & 22    & 18    & 26    & 22    & 23    & 18    & 20    & 20    & 26    & 22 & 27    & 23 & \textbf{\circledfill{28}} & \textbf{\squared{25}}  \\

          \hline
    \multirow{2}[0]{*}{0.7} & 20    & 27    & 22    & 32    & 27    & 29    & 23    & 25    & 24    & 34 & 27 & 34 & 29 & \textbf{\circledfill{35}} & \textbf{\squared{31}}  \\
          & 30    & 39    & 32    & 46    & 40    & 42    & 35    & 35    & 36    & 47    & 41 & 47    & 42 & \textbf{\circledfill{49}} & \textbf{\squared{45}}  \\

          \hline
    \multirow{2}[0]{*}{0.9} & 20    & 42    & 35    & 50    & 43    & 46    & 37    & 40    & 39    & 53 & 44 & 52    & 45 & \textbf{\circledfill{54}} & \textbf{\squared{48}}  \\
          & 30    & 59    & 50    & 68    & 62    & 65    & 56    & 57    & 58    & 70    & 64 & 70 & 65 & \textbf{\circledfill{71}} & \textbf{\squared{66}}  \\
\hline    \end{tabular}%
}
\label{tab5}
\end{table}%


Consider the powers against the exponential mixture, shown in Table \ref{tab4}. When using the MLE, the best performing test is $MP_n^{(2)}$, very closely followed by $ZA_{n}$ and $CV_n$. 
When turning our attention to the results obtained using the MME, the $MP_n^{(2)}$ test is nearly uniformly most powerful against the mixtures considered, but it is closely followed by $G_{n,1}$. Overall, the most powerful test is 
$MP_n^{(2)}$ using the MME.

The powers shown in Table \ref{tab5} indicate that the best performing test using the MLE is $MP_n^{(2)}$, very closely followed by $MP_n^{(1)}$. Turning our attention to the results pertaining to the MME, we see that $MP_n^{(2)}$ is not outperformed by any test against any of the mixtures considered. However, we again note that the $G_{n,1}$ test achieves powers close to those of $MP_n^{(2)}$.

The results above serve to support our findings obtained against fixed alternatives. That is, we recommend the use of the MME in order to perform parameter estimation, and we recommend using either the $MP_n^{(2)}$ or $G_{n,1}$ to test the hypothesis of the Pareto distribution.

\section{Practical application} \label{sect4}

We now use the tests developed for the Pareto distribution (i.e., we do not include the tests based on log-transformed data) to test whether or not two observed data sets are realised from the Pareto distribution. The data sets considered are the 2022 earnings of male golf players on the PGA (Professional Golfers' Association of America) and LIV Golf tours, respectively. LIV Golf is a professional golf tour started in Saudi Arabia and financed by their sovereign wealth oil fund. LIV represents the number 54 in Roman numerals and refers to both the number of holes played at each LIV tour event, as well as the number of shots played on a standard par 72 course if every hole played results in a birdie. The inaugural season started on 9 June 2022. To make the LIV Golf tour more enticing, the organisers offer prize earnings that are much higher than those offered by the more established PGA tour.

Table \ref{tab6} shows the earnings of the 28 LIV Golf tour players who earned more than \$3.5 million in the 2022 season. The average number of events played by these 26 players on the LIV Golf tour is only 7.85, whilst the average earnings is \$7 989 306 per player.
Table \ref{tab7} shows the earnings of the 28 PGA tour players who earned more than \$3.5 million in the 2022 season. The average number of events played by these 28 players on the PGA tour is 23.18, where this average decreases to 17.89 if one only counts the tournaments in which a player made the cut (i.e., played on the final two days of the tournament). The average earnings is \$6 098 395 per player and the average player earnings per tournament (based on 17.89 events) is \$340 882.90. From these summary statistics, as well as the entries in Tables \ref{tab6} and \ref{tab7}, it is clear that the earnings for the top earners are much higher for those playing in the LIV tour compared to those playing in the PGA tour.

\begin{table}
\begin{centering}
\begin{tabular}{cccccccccc}
\hline 
\$36 071 517 & \$16 993 416 & \$15 124 499 & \$13422785 & \$12 765 714 & \$9 792 500
& \$8 755 785\\
\$8 297 000 & \$8169167 & \$8 033 500 & \$7 638 000 & \$6 755 314 & \$5 741 000
& \$5 718 500\\
\$5 109 000 & \$4 992 618 & \$4 843 367 & \$4 614 500 & \$4 596 000 & \$4 535 000
& \$4 459 964\\
\$4 434 314 & \$4 382 417 & \$3 877 583 & \$3 700 000 & \$3 693 666 & \$3 599 100 &\$3 584 333\\
\hline 
\end{tabular}
\par\end{centering}
\caption{Season earnings of LIV golf players earning in excess of \$3.5 million.}
\label{tab6}
\end{table}

\begin{table}
\begin{centering}
\begin{tabular}{cccccccccc}
\hline 
\$14 046 909 & \$10 107 897 & \$9 405 081 & \$9 369 605 & \$8 654 566 & \$7 427 299 & \$7 073 986\\
\$7 012 672 & \$6 829 575 & \$6 520 597 & \$6 117 886 & \$5 776 298 & \$5 567 974
& \$5 289 842\\
\$5 248 220 & \$5 076 060 & \$5 018 443 & \$4 940 600 & \$4 868 461 & \$4 837 271
& \$4 722 433\\
\$4 310 047 & \$3 940 513 & \$3 876 590 & \$3 757 425 & \$3 718 990 & \$3 623 137
& \$3 616 679\\
\hline
\end{tabular}
\par\end{centering}
\caption{Season earnings of PGA players earning in excess of \$3.5 million.}
\label{tab7}
\end{table}

\begin{figure}[htp]
    \centering
    \includegraphics[width=12cm]{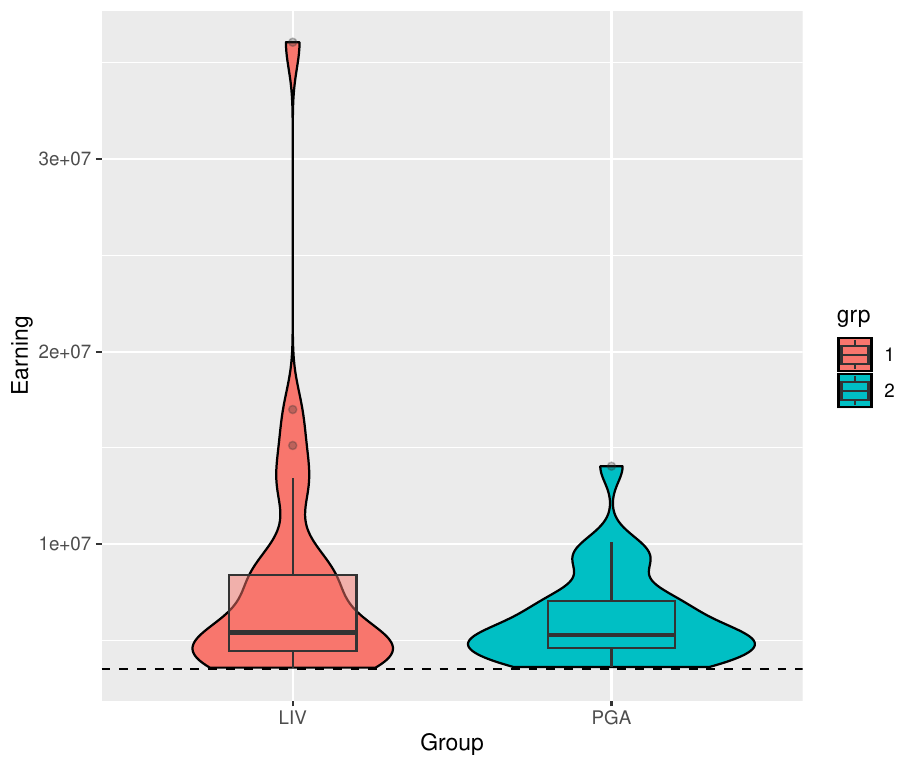}
    \caption{Violin and box plots for PGA and LIV golf earnings above \$3.5 millions.}
    \label{fig1}
\end{figure}

Figure \ref{fig1} shows side-by-side violin plots with overlayed box plots for the earnings data given in Tables \ref{tab6} and \ref{tab7}. Both density estimates indicate that the distributions of the earnings are positively skewed. Furthermore, both plots show evidence of the presence of outliers.

We test the hypotheses that the earnings of the top players of the PGA tour and the LIV Golf tour, respectively, are realised from a Pareto distribution. Tables \ref{tab8} and \ref{tab9} show the values of the test statistics as well as the corresponding $p$-values for the PGA and LIV Golf earnings, respectively. In each case we use both the MLE and the MME for estimation.

\begin{table}
\begin{centering}
\begin{tabular}{|l|ll|ll|}
\hline 
& MME && MLE & \\
Test & Statistic  & $p$-value &  Statistic  & $p$-value \tabularnewline
\hline 
$KS_n$ & 0.255 & 0.0243 & 0.206 & 0.1284 \\
$CV_n$ & 0.356 & 0.0505 & 0.177 & 0.2668 \\
$AD_n$ & 1.655 & 0.0741 & 0.891 & 0.3313 \\
$ZA_n$ & 3.484 & 0.1302 & 3.440 & 0.2273 \\
$MP^{(1)}_n$ & 0.009 & 0.0489 & 0.005 & 0.2849 \\
$MP^{(2)}_n$ & 0.009 & 0.0448 & 0.004 & 0.3135 \\
$G_{n,1}$ & 0.225 & 0.0412 & 0.045 & 0.4480 \\
\hline 
\end{tabular}
\par\end{centering}
\caption{Summary results for the PGA results and ranking.}
\label{tab8}
\end{table}

\begin{table}
\begin{centering}
\begin{tabular}{|l|ll|ll|}
\hline
& MME && MLE & \\
Test & Statistic  & $p$-value &  Statistic  & $p$-value \tabularnewline
\hline 
$KS_n$ & 0.151 & 0.4865 & 0.119 & 0.7714 \\
$CV_n$ & 0.110 & 0.5154 & 0.047 & 0.8961 \\
$AD_n$ & 0.603 & 0.5926 & 0.284 & 0.9473 \\
$ZA_n$ & 3.337 & 0.7631 & 3.309 & 0.9584 \\
$MP^{(1)}_n$ & 0.003 & 0.4958 & 0.001 & 0.9212 \\
$MP^{(2)}_n$ & 0.003 & 0.4457 & 0.002 & 0.9174 \\
$G_{n,1}$ & 0.069 & 0.3791 & 0.004 & 0.8813 \\
\hline
\end{tabular}
\par\end{centering}
\caption{Summary results for the LIV Golf results and ranking.}
\label{tab9}
\end{table}

A striking feature of Table \ref{tab8} is the substantial difference between the $p$-values obtained using the two estimation methods. 
The results indicate that the null hypothesis of the Pareto distribution is rejected by $KS_n$, $MP_n^{(1)}$, $MP_n^{(2)}$ and $G_{n,1}$ at a 5\% significance level when using the MME to estimate $\beta$, while $CV_n$ fails to reject the null hypothesis in this case by the smallest of margins. 
If we set the nominal significance level to 10\%, then all of the tests considered, with the exception of $ZA_n$, reject the null hypothesis. 
On the other hand, none of the tests reject the hypothesis of the Pareto distribution at the 10\% level when using the MLE. 
In the Monte Carlo power study, we recommended using $MP_n^{(2)}$ or $G_{n,1}$ together with the MME. As a result, the rejection of these tests using the MME casts serious doubt as to the hypothesis that the earnings associated with the PGA tour are realised from a Pareto distribution.

We now turn our attention to the 
LIV Golf results shown in Table \ref{tab9}. In this case, all seven tests considered fail to reject the null hypothesis for any reasonable level of significance, regardless of how $\beta$ is estimated. We conclude that there is little doubt as to the compatibility of these data with the Pareto assumption. 
One possible reason for the discrepancy in the results of Tables \ref{tab8} and \ref{tab9} is the higher prize money associated with LIV Golf compared to the PGA tour. The heavy tail of the Pareto seems to provide a reasonable model for these extreme earnings.

\section{Conclusions} \label{sect5}

We propose two new goodness-of-fit tests for the Pareto distribution based on a multiplicative version of the memoryless property which characterises this distribution. The resulting tests are computationally inexpensive as both test statistics require only the calculation of single summations.

A Monte Carlo study investigates the empirical powers associated with the new tests and compares these results to those of existing tests in the literature. 
We consider powers against fixed alternatives as well as mixture distributions. In both cases, the newly proposed tests are found to be competitive, often outperforming the existing tests against the alternative distributions considered.

A result of independent interest is to compare the powers associated with tests developed specifically for the Pareto distribution to that of tests developed for the exponential distribution when applied to log-transformed data. The simulation results indicate that the latter underperform relative to the former by some margin. 

A practical example relating to the earnings of golfers is included. The seasons' earnings of golfers earning in excess of \$3.5 million in both the PGA and LIV Golf are considered separately. In each case, we test the hypothesis that the observed earnings are realised from a Pareto distribution. Interestingly, we find that the Pareto distribution provides an accurate model for the earnings from LIV Golf, while there is evidence suggesting that the earnings of the PGA are not realised from this distribution.

\appendix
\section*{Appendix}

Table \ref{tab10} and \ref{tab11} show the empirical powers associated with the exponentiality tests based on the log-transformed data. For ease of comparison, the two highest powers (including ties) against each alternative distribution are indicated in bold.

\begin{table}[htbp]
  \centering
  \caption{Numerical powers of tests based on log-transformed data with $n=20$} \label{Exp20}
    \begin{tabular}{|l|r|r|r|r|r|r|r|}
          \hline
      Distribution    & {$\widetilde{KS}_n$} & {$\widetilde{CV}_n$} & {$\widetilde{AD}_n$} & {$\widetilde{ZA}_n$} & {$\widetilde{AG}_n$} & {$\widetilde{CO}_n$} & {$\widetilde{BH}_n$} \\
      \hline
    P(2)  & \textbf{5} & \textbf{5} & \textbf{5} & \textbf{5} & \textbf{5} & \textbf{5} & \textbf{5} \\
    P(5)  & \textbf{5} & \textbf{5} & \textbf{5} & \textbf{5} & \textbf{5} & \textbf{5} & \textbf{5} \\
    P(10) & \textbf{5} & \textbf{5} & \textbf{5} & \textbf{5} & \textbf{5} & \textbf{5} & \textbf{5} \\
    $\Gamma(0.8)$ & 10    & \textbf{11} & 10    & \textbf{11} & 10    & \textbf{13} & \textbf{11} \\
    $\Gamma(1)$ & 25    & 31    & 26    & 28    & 25    & \textbf{39} & \textbf{33} \\
    $\Gamma(1.2)$ & 47    & 58    & 53    & 56    & 46    & \textbf{69} & \textbf{62} \\
    W(0.8) & 8     & \textbf{9} & \textbf{9} & \textbf{9} & 8     & \textbf{9} & 8 \\
    W(1.2) & 51    & 62    & 57    & 60    & 49    & \textbf{73} & \textbf{66} \\
    W(1.5) & 82    & 92    & 90    & 91    & 83    & \textbf{96} & \textbf{94} \\
    LN(1) & 55    & 66    & 65    & \textbf{80} & 52    & \textbf{88} & 66 \\
    LN(1.5) & 7     & 7     & 6     & \textbf{10} & 9     & \textbf{14} & 7 \\
    LN(2.5) & 26    & 29 & \textbf{43} & 26    & 10    & \textbf{47}     & 25 \\
    HN(0.5) & 47    & 59    & 54    & 54    & 48    & \textbf{62} & \textbf{63} \\
    HN(1) & 53    & 65    & 60    & 60    & 53    & \textbf{68} & \textbf{69} \\
    HN(1.5) & 62    & 71    & 69    & 68    & 61    & \textbf{74} & \textbf{76} \\
    LFR(0.2) & 33    & 41    & 35    & 37    & 32    & \textbf{47} & \textbf{44} \\
    LFR(0.8) & 42    & 53    & 47    & 48    & 43    & \textbf{58} & \textbf{57} \\
    LFR(1) & 44    & 56    & 50    & 51    & 45    & \textbf{61} & \textbf{60} \\
    BE(0.8) & 11    & 12    & 11    & 12    & 11    & \textbf{13} & \textbf{13} \\
    BE(1) & 24    & 31    & 26    & 28    & 24    & \textbf{39} & \textbf{33} \\
    BE(1.5) & 67    & 80    & 77    & 80    & 67    & \textbf{89} & \textbf{83} \\
    TP(1) & 12    & \textbf{13} & 11    & \textbf{13} & \textbf{13} & \textbf{19} & \textbf{13} \\
    TP(2) & 21    & \textbf{25} & 21    & 24    & 22    & \textbf{34} & \textbf{25} \\
    TP(3) & 31    & \textbf{38} & 32    & 35    & 31    & \textbf{47} & \textbf{38} \\
    D(0.4) & 28    & 34    & 30    & \textbf{38} & 28    & \textbf{52} & 35 \\
    D(0.6) & 50    & 61    & 57    & \textbf{65} & 48    & \textbf{80} & 63 \\
    D(0.8) & 71    & 82    & 79    & \textbf{84} & 69    & \textbf{93} & \textbf{84} \\
    \hline
    \end{tabular}%
  \label{tab10}
\end{table}%

\begin{table}[htbp]
  \centering
  \caption{Numerical powers of tests based on log-transformed data with $n=30$} \label{Exp30}
    \begin{tabular}{|l|r|r|r|r|r|r|r|}
    \hline
           Distribution    & {$\widetilde{KS}_n$} & {$\widetilde{CV}_n$} & {$\widetilde{AD}_n$} & {$\widetilde{ZA}_n$} & {$\widetilde{AG}_n$} & {$\widetilde{CO}_n$} & {$\widetilde{BH}_n$} \\
      \hline
    $P(2)$  & \textbf{5} & \textbf{5} & \textbf{5} & \textbf{5} & \textbf{5} & \textbf{5} & \textbf{5} \\
    $P(5)$  & \textbf{5} & \textbf{5} & \textbf{5} & \textbf{5} & \textbf{5} & \textbf{5} & \textbf{5} \\
    $P(10)$ & \textbf{5} & \textbf{5} & \textbf{5} & \textbf{5} & \textbf{5} & \textbf{5} & \textbf{5} \\
    $\Gamma(0.8)$ & 13    & \textbf{14} & 13    & \textbf{14} & 12    & \textbf{14} & \textbf{15} \\
    $\Gamma(1)$ & 36    & 45    & 40    & 44    & 33    & \textbf{53} & \textbf{49} \\
    $\Gamma(1.2)$ & 64    & 77    & 73    & 77    & 61    & \textbf{85} & \textbf{81} \\
    $W(0.8)$ & 10    & \textbf{11} & \textbf{11} & \textbf{12} & 10    & \textbf{11} & \textbf{11} \\
    $W(1.2)$ & 69    & 82    & 79    & 81    & 66    & \textbf{89} & \textbf{86} \\
    $W(1.5)$ & 96    & 99    & 99    & 99    & 96    & \textbf{100} & \textbf{100} \\
    $LN(1)$ & 76    & 85    & 87    & \textbf{96} & 72    & \textbf{98} & 86 \\
    $LN(1.5)$ & 8     & 9     & 8     & \textbf{15} & 10    & \textbf{20} & 9 \\
    $LN(2.5)$ & 39    & 43 & \textbf{58} & 39    & 24    & \textbf{61}     & 36 \\
    $HN(0.5)$ & 67    & 78    & 75    & 77    & 65    & \textbf{80} & \textbf{83} \\
    $HN(1)$ & 72    & \textbf{84} & 81    & 83    & 72    & \textbf{84} & \textbf{88} \\
    $HN(5.5)$ & 77    & 89    & 86    & 89    & 76    & \textbf{90} & \textbf{95} \\
    $LFR(0.2)$ & 46    & 56    & 51    & 56    & 43    & \textbf{62} & \textbf{61} \\
    $LFR(0.8)$ & 60    & 72    & 68    & 71    & 58    & \textbf{76} & \textbf{77} \\
    $LFR(1)$ & 64    & 77    & 72    & 74    & 61    & \textbf{79} & \textbf{81} \\
    $BE(0.8)$ & 14    & \textbf{17} & 15    & \textbf{17} & 14    & \textbf{17} & \textbf{17} \\
    $BE(1)$ & 35    & 44    & 39    & 42    & 32    & \textbf{52} & \textbf{49} \\
    $BE(1.5)$ & 85    & 94    & 93    & 95    & 84    & \textbf{98} & \textbf{96} \\
    $TP(1)$ & 15    & \textbf{17} & 14    & \textbf{17} & 16    & \textbf{25} & \textbf{17} \\
    $TP(2)$ & 32    & 37    & 32    & 35    & 29    & \textbf{47} & \textbf{38} \\
    $TP(3)$ & 45    & 53    & 48    & 49    & 42    & \textbf{63} & \textbf{55} \\
    $D(0.4)$ & 42    & 50    & 48    & \textbf{56} & 40    & \textbf{70} & 52 \\
    $D(0.6)$ & 71    & 81    & 80    & \textbf{85} & 67    & \textbf{94} & 84 \\
    $D(0.8)$ & 89    & 95    & 95    & \textbf{96} & 86    & \textbf{99} & \textbf{96} \\
    \hline
    \end{tabular}%
  \label{tab11}
\end{table}%
\newpage

\bibliographystyle{apa-good}
\bibliography{Ref}

\end{document}